\documentclass{emulateapj}
\usepackage{amsmath}
\usepackage{amssymb}
\usepackage{apjfonts}
\usepackage{epsfig}
\usepackage{natbib}



\newcommand{\kms}{\,{\rm km\,s^{-1}}}
\newcommand{\au}{\,{\rm AU}}

\newcommand{\yr}{\,{\rm yr}}
\newcommand{\myr}{\,{\rm Myr}}

\newcommand{\pc}{\,{\rm pc}}

\newcommand{\ergs}{\,{\rm erg\,s^{-1}}}

\newcommand{\msun}{\,M_\odot}
\newcommand{\rsun}{\,R_\odot}

\newcommand{\be}{\begin{equation}}
\newcommand{\ee}{\end{equation}}

\newcommand{\bea}{\begin{eqnarray}}
\newcommand{\eea}{\end{eqnarray}}

\newcommand{\ben}{\begin{enumerate}}
\newcommand{\een}{\end{enumerate}}

\newcommand{\mbh}{M_\bullet} 
\newcommand{\mb}{M_b}

\newcommand{\calr}{\mathcal{R}}

\newcommand{\vsp}{\vspace{3mm}}
\newcommand{\vspb}{\vspace{1pt}}

\begin{document}

\shorttitle{BINARY DISRUPTION}
\shortauthors{PFAHL}


\submitted{Accepted by ApJ Letters}

\title{Binary Disruption by Massive Black Holes in Globular Clusters}

\author{Eric Pfahl}

\affil{Chandra Fellow; Department of Astronomy, University of
Virginia, P.O. Box 3818, Charlottesville, VA 22903-0818;
epfahl@virginia.edu}


\begin{abstract}
  
A massive ($\sim$$10^2$--$10^4\msun$) black hole at the center of a
globular cluster will tidally disrupt binaries that pass sufficiently
close.  Such an encounter results in the capture of one binary
component by the black hole and ejection of the other star.  A
$10^3\msun$ black hole disrupts one binary every 1--10\,Myr; captured
stars orbit with initial periods of days to decades.  Although
$\ga$100 stars are captured in only 1\,Gyr, the number with periods of
$<$10\,yr at any one time is probably less than 10, regulated by
stellar interactions and tidal disruption by the black hole.  Binary
disruption and stellar capture should be critically examined with
detailed numerical simulations, in order to corroborate the analytic
results given here and to weigh possible observational consequences.

\end{abstract}




\section{INTRODUCTION}\label{sec:intro}\vspb

Black holes of mass $\sim$$10^2$--$10^4\msun$ occupy a special niche
in several subfields of theoretical astrophysics.  Such objects might
naturally form as remnants of the first stars in the Universe
\citep[][]{barkana01,madau01,schneider02}, or be built through merging
of stars \citep{zwart02} or stellar-mass black holes \citep{miller02}
in young star clusters.  These black holes may play critical roles in
the formation and growth of supermassive ($\ga$$10^6\msun$) black
holes in galactic nuclei \citep[][]{ebisuzaki01,islam04}.

Intermediate-mass black holes (IMBHs) have been invoked to resolve
several recent observational quandaries.  Some of the
``ultraluminous'' X-ray sources detected in other galaxies may be
accreting IMBHs (e.g, Miller \& Colbert 2004), although the majority
are likely stellar-mass black holes \citep[][]{king01,rappaport04}.
There have been recent claims of possible IMBH detections in two
globular clusters (M15 in the Milky Way and G1 in M31) based on {\em
HST} data \citep{gebhardt02,gerssen02}.  However, the data do not
require IMBHs, as shown by \citet{baumgardt03a,baumgardt03b} and
\citet{mcnamara03}.  \citet{colpi02} cite evidence for a single or
binary IMBH in the cluster NGC~6752 based on the anomalous positions
and accelerations of cluster radio pulsars.  Finally, a single or
binary IMBH could account for the net rotation observed in the center
of M15 \citep{gebhardt00,gerssen02,miller04}.
 
Because of the broad scientific scope sampled by IMBHs, it is crucial
to investigate what unique dynamical signatures they imprint on their
environments.  The high stellar densities and proximities of Galactic
globular clusters make them prime locales to search for IMBHs, as
emphasized by much of the recent observational work; for a review of
early theoretical and observational studies, see \citet{shapiro85}.
Here we introduce a process that deposits stars directly into compact
orbits (e.g., $\la$100\,AU) around a massive cluster black hole.

Binary stars are abundant in globular clusters \citep[e.g.,][]{hut92}.
A binary that passes sufficiently near a central black hole is
disrupted by tidal forces, resulting in the capture of one star and
ejection of the other (see \S~\ref{sec:dyn}).  This type of
interaction was studied in the context of Sgr A$^*$ by \citet{hills88}
and \citet{yu03} as a mechanism for accelerating stars to very high
speeds, and by \citet{gould03} to account for the young stars near Sgr
A$^*$ \citep[e.g.,][]{eisenhauer05}.  The disruption rate in globular
clusters is estimated in \S~\ref{sec:binlc}, and the fate of captured
stars is discussed in \S~\ref{sec:clusdyn}.  Future directions for
studying this mechanism in more detail are suggested in
\S~\ref{sec:dis}, as are potentially observable consequences.


\section{Binary Disruption}\label{sec:dyn}

Consider a binary system of mass $M_b = M_1 + M_2$ with semimajor axis
$a_b$ on a parabolic approach to a black hole of mass $\mbh \gg M_b$.
Let $r$ be the distance between the binary barycenter and the black
hole, and $s$ be the {\em minimum} distance.  The differential
acceleration across the binary ($\sim$$G\mbh a_b/r^3$) exceeds the
binary self-gravity ($\sim$$GM_b/a_b^2$) within the tidal radius,
\vsp
\begin{equation}\label{eq:tidrad}
r_t \equiv a_b\,(\mbh/M_b)^{1/3} 
= 10\,a_b Q_3^{1/3}~, \vsp
\end{equation}
where $\mbh/\mb = 10^3 Q_3$.  For a disrupted binary, the tidal
perturbation is essentially impulsive, and the additional relative
velocity imparted to the binary components has a magnitude of
$\sim$$v_b = (GM_b/a_b)^{1/2}$.  Immediately following the disruption,
the two stars are mutually unbound and move relative to the black hole
with the original systemic velocity of magnitude $u_t \simeq
(2G\mbh/r_t)^{1/2}$, plus contributions from the prior orbital motion
and tidal perturbation.  Random orbital orientations and phases at the
time of disruption lead to a range of orbital parameters and
asymptotic speeds for stars that are captured or ejected; below are
estimates for the typical values.


If star 1 is captured, we can estimate the expectation value of its
specific energy, $E_1$, relative to the black hole by assuming that
its resultant velocity after the tidal perturbation is directed
opposite to the binary systemic velocity and has a magnitude of
$(M_2/M_b)v_b$.  To leading order, we obtain \vsp
\begin{equation}\vsp
E_1 \simeq \frac{1}{2}\left(u_t - \frac{M_2}{M_b}v_b\right)^2 
- \frac{G\mbh}{r_t}
\simeq -\frac{M_2}{M_b}u_tv_b~. \vsp
\end{equation}
The corresponding semimajor axis and orbital
period are
\vsp
\begin{equation}\label{eq:semicap}
a_1 = -G\mbh/2E_1 
\simeq 35\,a_b\,(M_b/M_2)Q_3^{2/3}~, \vsp
\end{equation}
and
\vsp
\begin{equation}\label{eq:porbcap}
P_1 \simeq 
7\,P_b\,(M_b/M_2)^{3/2}Q_3^{1/2}~, \vsp
\end{equation}
where $P_b$ is the binary period.  The pericenter separation of star
1, $a_1(1 - e_1)$, must be comparable to $s$, so that
\vsp
\begin{equation}\label{eq:ecccap}
e_1 \simeq 1 - \frac{s}{a_1}
\simeq 
1 - 0.3\left(\frac{s}{r_t}\right)
\left(\frac{M_2}{M_b}\right)Q_3^{-1/3}~, \vsp
\end{equation}
which is $\ga$0.7 for $Q_3 > 1$.  Star 2 has a typical specific energy
of $E_2\simeq (M_1/M_b)u_tv_b$, and thus its speed at infinity is
\vsp
\begin{equation}
v_2 \simeq 4 \,v_b\,(M_1/M_b)^{1/2}Q_3^{1/6}~, \vsp
\end{equation}
which may be $\ga$$100\kms$ for typical binary parameters.  A star
ejected by this process can easily escape its host cluster.


\section{Total Disruption Rate}\label{sec:binlc}

We now employ ``loss-cone'' theory to estimate the rate of binary
disruptions.  The approach taken here is not very rigorous, but
captures the essential physics.  More precise accounts of loss-cone
dynamics can be found in, e.g., \citet{frank76}, \citet{lightman77},
\citet{cohn78}, \citet{syer99}, and \citet{magorrian99}.

A binary with semimajor axis $a_b$ at radius $r \gg r_t$ is tidally
disrupted if its specific angular momentum about the black hole is
less than $J_t = (2G\mbh r_t)^{1/2}$.  If the local one-dimensional
stellar velocity dispersion, $\sigma(r)$, is adopted as a typical
binary systemic speed, the disruption condition defines a double-sided
loss cone in velocity space with semiangle $\theta_t = J_t/r\sigma$.
Perturbations by distant stars cause the binary barycenter to undergo
a random walk in phase space.  In a dynamical time, $\tau_d \sim
r/\sigma$, the velocity is deflected by an angle $\theta_d \sim
(\tau_d/\tau_r)^{1/2}$, measured at radius $r$.  The local two-body
relaxation timescale is given by \citep[e.g.,][]{spitzer71}
\vsp
\begin{equation}\label{eq:relax}
\tau_r(r) = \frac{\sigma^3(r)}{\Lambda G^2 m^2 n(r)}
\simeq 50\,\frac{(\sigma/10\kms)^3}{\Lambda_{10}\,(n/10^5\pc^{-3})}\myr~, \vsp
\end{equation}
where $n(r)$ is the stellar number density, $\Lambda =
\Lambda_{10}\,10$ is proportional to the Coulomb logarithm, and we
have assumed $m = 1\msun$ for the mean stellar and binary mass, which
is fixed in the subsequent calculations.

The ratio $\theta_d/\theta_t$ increases monotonically with $r$ for
profiles $n(r)$ and $\sigma(r)$ suitable to globular clusters.  Two
loss-cone regimes are loosely demarcated by $\theta_d/\theta_{\rm lc}
= 1$, defining a critical radius $r_c$.  For $r > r_c$, binaries can
scatter into and out of the loss cone in a dynamical time, and the
loss cone is full statistically.  If $r < r_c$, binaries within the
loss cone are disrupted in a dynamical time, and the disruption rate
is limited by the diffusion of binaries across the loss-cone boundary.

We assume that the cluster center consists of a power-law density cusp
around the black hole, embedded in an isothermal cluster core of
radius $r_0$.  The cusp is inside the radius of influence of the
central black hole: \vsp
\begin{equation}
r_i = G\mbh/\sigma_i^2 
\simeq 0.04\,M_3\sigma_{10}^{-2}\pc~, \vsp
\end{equation}
where $\mbh = 10^3 M_3\msun$, and $\sigma_i = 10\sigma_{10}\kms$ is
the velocity dispersion at $r_i$.  Each zone has different power-law
profiles, $n(r) = n_i(r/r_i)^{-\alpha}$ and $\sigma^2(r) =
\sigma_i^2(r/r_i)^{-\beta}$, where $\alpha = \beta = 0$ in the core,
and $\alpha \simeq 1.5$--2 \citep[e.g.,][]{bahcall77,young77} and
$\beta = 1$ in the cusp. The critical radius is \vsp
\begin{equation}\label{eq:rcrit}
r_c/r_i = (a_b/a_c)^{1/(3+\beta - \alpha)}~, \vsp
\end{equation}
where, given $n_i = 10^5 n_5 \pc^{-3}$, we have 
\vsp
\begin{equation}
a_c \simeq 0.1\, n_5\,\,M_3^{5/3}\,\sigma_{10}^{-8}\Lambda_{10}\au~. \vsp
\end{equation}
%


Cluster binaries interact frequently with single stars and other
binaries.  These encounters shape the binary population as the cluster
and its stars evolve, a complex problem that can only be fully
unraveled with detailed cluster simulations
\citep[e.g.,][]{fregeau03,giersz03}.  Here we assume for simplicity a
differential distribution $p(a_b) \propto a_b^{-1}$ over the range
$a_b = 0.01$--1\,AU, and a fixed binary number fraction $f_b$.  As the
cluster evolves, the shape and limits of $p(a_b)$ vary, and $f_b$ in
the core may decrease from $\ga$$0.5$ initially to current values of
$\la$$0.1$ \citep[][]{ivanova05}.

When the loss cone is full, a fraction $\simeq$$\theta_t^2$ of the
binaries with semimajor axis $a_b$ at radius $r$ are disrupted in a
dynamical time. The differential rate is \vsp
\begin{equation}\label{eq:lcfull}
\frac{d\calr_{\rm full}}{d a_b} \simeq
4\pi f_b G \mbh \left(\frac{\mbh}{m}\right)^{1/3}
p(a_b)a_b 
\int_{r_c}^{r_0}\frac{dr}{r}\frac{n(r)}{\sigma(r)}~. \vsp
\end{equation}
The contribution from $r > r_i$ dominates, and we obtain
\vsp
\begin{equation}\label{eq:fullnum}
\calr_{\rm full} \sim 
10^{-5}\,f_b n_5\, \langle a_b({\rm AU})\rangle \, 
M_3^{4/3} \sigma_{10}^{-1} \ln\left(\frac{r_0}{r_i}\right)
\yr^{-1}~, \vsp
\end{equation}
where $\langle a_b({\rm AU})\rangle$ is the mean semimajor axis in AU.

In the diffusive regime, the fraction of binaries disrupted in a time
$\tau_r(r)$ is $\sim$$1/\ln(2/\theta_t)$ \citep[e.g.,][]{frank76}.
The differential diffusive disruption rate is then
\vsp
\begin{equation}\label{eq:lcdiff}
\frac{d\calr_{\rm diff}}{d a_b} \simeq
4\pi f_b G^2 m^2
p(a_b) \int^{r_c}\frac{dr}{r} \frac{r^3n^2(r)}{\sigma^3(r)}~, \vsp
\end{equation}
where the weakly varying factor $\Lambda/\ln(2/\theta_t) \sim$1 has
been dropped.  The lower limit in the above radial integral is
relatively unimportant if it is $\ll$$r_c$. Numerically, we find
\vsp
\begin{equation}\label{eq:diffnum}
\calr_{\rm diff}  \sim 
10^{-6}\,f_b n_5^2 \,M_3^3 \,\sigma_{10}^{-9}\yr^{-1}~. \vsp
\end{equation}

A binary fraction of $f_b \sim 0.1$ and mean semimajor axis of
$\langle a_b \rangle \sim 0.1\au$ give disruption rates of
$\sim$$10^{-7}\yr^{-1}$ for both the full and diffusive regimes when
$M_3 = n_5 = \sigma_{10} =1$.  The total disruption rate is then
$\sim$$10^{-7}$--$10^{-6}\yr^{-1}$ for the same parameter values.  For
$a_b = 0.01$--1\,AU and $M_3 \simeq 1$, captured stars have $a_1 \sim
1$--100\,AU when $M_1 \simeq M_2$, and periods of $\sim$10\,d to
$\sim$$30\yr$.  From eq.~(\ref{eq:semicap}), $p(a_1)$ should have a
similar shape as $p(a_b)$.  Our choice of $p(a_b)\propto a_b^{-1}$
implies that the distributions of $\log a_1$, $\log P_1$, and $\log
v_2$ should be roughly flat.

As an illustration, we computed $10^4$ scattering encounters using a
binary-single scattering code similar in logic to that described in
\citet{hut83}.  The motion was integrated with the symplectic
algorithm of \citet{mikkola99} and \citet{preto99}.  We chose a
black-hole mass of $1000\msun$ and stellar masses of $M_1 = M_2 =
1\msun$.  The binary orbits were initially circular, with separations
drawn from $p(a_b) \propto a_b^{-1}$ ($a_b = 0.01$--1\,AU).  The speed
at infinity was fixed at $v_\infty = 10\kms$.  Impact parameters, $b$,
were drawn from a distribution that is flat in $b^2 \simeq 2sG(\mbh +
M_b)/v_\infty^2$, limited so that $s < 2r_t$.  Figure 1 shows that the
distributions of $\log a_1$ and $\log v_2$ are roughly flat in the
middle, with high and low tails not predicted in \S~\ref{sec:dyn}.
Solar-type stars with $a_1(1- e_1) < 10\rsun$ (below the dashed curve
in Fig.~1) are probably tidally disrupted by the black hole (see
eq. [\ref{eq:tidrad}]).

\begin{figure}
\centerline{\epsfig{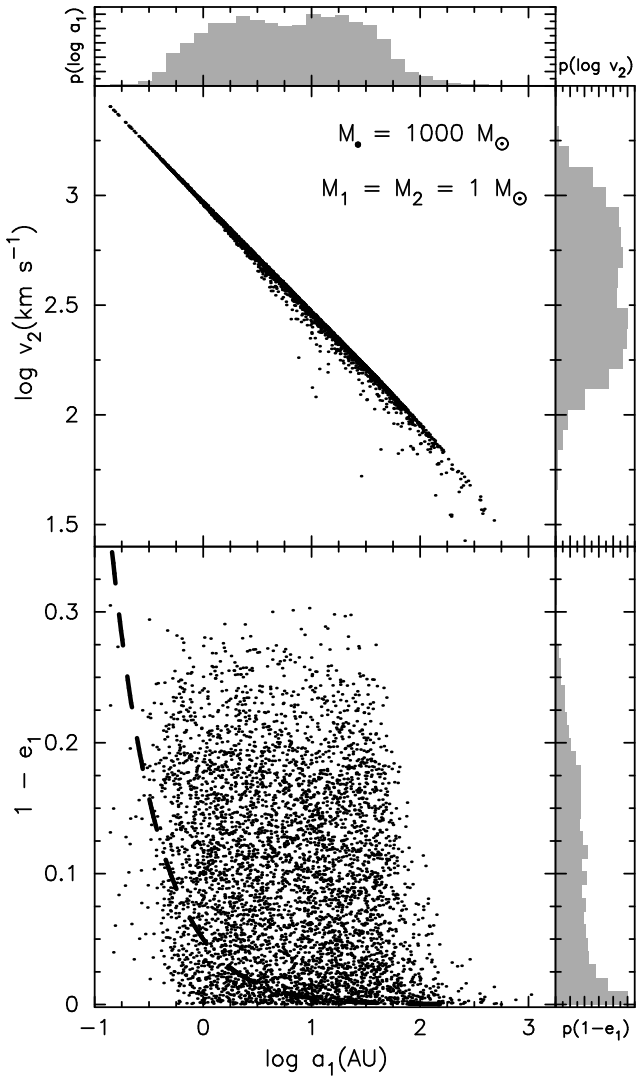}}
\caption{Simulated outcomes for binaries with $M_1 = M_2 = 1\msun$
disrupted by a $1000\msun$ black hole.  The panels to the top and
right of the scatter plots show the histograms of $\log a_1$, $\log
v_2$, and $1-e_1$.  Below the dashed curve ($a_1(1 - e_1) = 10\rsun$)
a solar-type star is tidally disrupted by the black hole.
\label{fig:cappar}}
\end{figure}
%


\section{Dynamical Evolution of Captured Stars}\label{sec:clusdyn}

Each binary disruption by a $\simeq$$10^3\msun$ black hole deposits
one star into a compact, eccentric Keplerian orbit. In 100\,Myr,
$\sim$10--100 stars may be captured.  However, a collection of $\ga$10
stars orbiting the black hole with periods of $\la$10\,yr will not
persist in the time between captures.  Mechanisms that drive the
evolution of captured stars include gravitational and collisional
two-body encounters, and tidal disruption by the black hole.  These
processes are quantified below, under the ad hoc simplifying
assumptions that the stars have equal masses of $m = 1\msun$, follow a
power-law density profile with $\alpha \simeq 2$, and everywhere have
a nearly isotropic velocity field.  We neglect relatively weak and
rare perturbative encounters between captured stars and cluster stars
on highly elongated orbits \citep[][]{gultekin04}.

Consider $N$ point masses orbiting the black hole.  Let $r_h$ be the
radius that contains half of the stars and $P_h$ be the period of a
circular orbit at $r_h$.  The relaxation time at $r_h$ is $\tau_h \sim
10^5 P_h\,M_3^2\,N^{-1}$ (eq.~[\ref{eq:relax}] with $\Lambda =
\ln(\mbh/m) \sim 10$).  ``Resonant relaxation'' \citep[][]{rauch96} of
a star's orbital angular momentum occurs on a shorter timescale of
$\tau_h^{\rm res} \sim 10^3\,P_h\,M_3$, driving the collection of
captured stars toward velocity isotropy without affecting the energy
distribution.  If $P_h \sim 10\yr$ and $N \ga 10$, the orbits spread
significantly in the radial direction in the time between captures.
Captured stars on the widest orbits will mingle with other
(non-captured) stars within $r_i$.  The outward transport of captured
stars into the cusp may noticeably alter the cusp structure.

A star may be unbound from the black hole following a single strong
encounter \citep[e.g.,][]{lin80}. An upper limit to the ejection rate
is obtained by setting the maximum two-body impact parameter to
$b_{\rm ej} = Gm/\sigma^2(r)$, characteristic of a $\sim$$90^\circ$
deflection but not necessarily an ejection.  The local rate per star
is $\sim$$\pi n(r)b^2_{\rm ej}\sigma(r)$.  Integrating over all
captured stars, we find a rate of
$\la$$10^{-7}\,P_h^{-1}\,M_3^{-2}\,N^2$.  For $N\sim 10$ and $P_h \sim
10\yr$, most stars may be ejected in the time between captures.  Under
the point-mass assumption, energy conservation demands that ejection
and relaxation leave at least one star tightly bound to the black
hole.  When finite stellar radii are considered, two additional
processes must be addressed.
 
First, the (non-resonant) rate at which the black hole tidally
disrupts captured stars is $\sim$$N/\tau_h \sim
10^{-5}\,P_h^{-1}\,M_3^{-2}\,N^2$, which is much more efficient than
ejection at removing stars.  Resonant relaxation may increase the rate
substantially, since angular momentum diffusion drives stars into the
loss cone.  Note that if a considerable fraction (e.g., $\ga$10\%) of
the captured stars are ultimately disrupted by the hole, the implied
rate may exceed the swallowing rate for single stars determined by
previous authors \citep[$\sim$$10^{-8}\yr^{-1}$ for $M_3 = n_5 =
\sigma_{10} =1$; e.g.,][]{frank76,lightman77}.  Second, the collision
rate for stars of radius $R$ is $\sim$$2\pi P_h^{-1}(R/r_h)^2 N^2 \sim
(10^{-9}$--$10^{-6})(R/R_\odot)^2N^2\yr^{-1}$ for $r_h = 10$--50\,AU.
A collision between two main-sequence stars will result in coalescence
unless the relative speed at infinity is significantly larger than the
escape speed from the stellar surface \citep[e.g.,][]{freitag05},
which may occur within $\sim$1\,AU of a $10^3\msun$ black hole.

When $N \ga 10$ and $P_h \sim 10\yr$, all of the above processes act
on timescales shorter than or comparable to the time between stellar
captures.  Thus, at any time, only a few stars are expected to orbit
the black hole with periods of $\la$10\,yr.



\section{Conclusions and Outlook}\label{sec:dis}

A $10^3\msun$ black hole at the center of a globular cluster may
tidally disrupt {\em thousands} of binaries over 10\,Gyr, each
encounter leaving one star within $\sim$100\,AU of the hole.  At any
given time, the number of stars in such compact orbits is probably
only a few, limited by interactions between captured stars and tidal
disruption.  Before we can assess possible observational consequences
of these processes, numerical simulations are required to properly
calculate the binary disruption rate and the dynamical evolution of
captured stars.

The cluster evolution codes of \citet[][Monte Carlo]{fregeau03} and
\citet[][gas-dynamical/Monte Carlo]{giersz03} are well suited to the
problem of determining the binary disruption rate.  These codes
efficiently treat a realistic number ($\ga$$10^5$) of stars and
binaries, and include explicit integration of close binary-single and
binary-binary encounters.  The black hole would be represented by a
central point mass, as in \citet[][]{freitag02}, but the imposed
spherical symmetry demands that it be fixed at the origin, neglecting
its Brownian motion.  These algorithms cannot adequately follow the
dynamics of a small-$N$ system of captured stars.

Direct $N$-body integration has been used to study the dynamics of
single stars near a massive black hole
\citep{baumgardt04b,preto04}. Such simulations are prohibitively
time-consuming when $N \ga 10^5$ and there is a sizable binary
fraction.  Simulations with $N \sim 10^4$ may be of value, since the
dynamics of captured stars are computed correctly.  It may also be
useful to follow the collisional $N$-body evolution $\sim$10 captured
stars in isolation from the cluster.

The high binary disruption rate and compact orbits of captured stars
increase the likelihood that a cluster IMBH will be detected, perhaps
in one of the following ways:

1. If, following a stellar tidal disruption, $\simeq$$0.5\msun$ is
accreted by the black hole with a 10\% mass-to-light conversion
efficiency, the duration of the Eddington-limited flare is
$\sim$$10^4M_3^{-1}\yr$.  Such a flare might be visible as a very
bright UV/X-ray source in one of the $\sim$100 globular clusters in
the Milky Way if most contain a $\simeq$$10^3\msun$ black hole that
disrupts stars at a rate of $\sim$$10^{-6}\yr^{-1}$; but we do not,
and both suppositions may be highly optimistic.

2. If the orbit of a captured star with a pericenter distance of
$\simeq$$10M_3^{1/3}\rsun$ circularizes without the star being
destroyed, stellar evolution may drive a long ($\ga$$10^8\yr$) phase
of mass transfer, with accretion rates of $\la$$10^{-8}\msun\yr^{-1}$
and luminosities of $\la$$10^{38}\ergs$ \cite[][]{hopman04,zwart04}.
Steady mass transfer from a $\ga$$10\msun$ donor has been proposed as
a fuel source for ultraluminous X-ray sources in young star clusters.

3. Hundreds of cluster neutron stars may acquire binary companions
dynamically, accrete mass, and ultimately turn on as millisecond radio
pulsars.  A $\simeq$$10^3\msun$ black hole may capture 1 pulsar from a
binary system every $\sim$1\,Gyr, making the detection of such a
pulsar unlikely.  The timing properties of an orbiting pulsar would
clearly reveal the black hole.

4. A captured neutron star or white dwarf that migrates to within
$\sim$$1\rsun$ of a $10^3\msun$ black hole would produce a detectable
gravitational-wave signal for {\em LISA} from distances of
$\la$100\,Mpc in the short time ($\la$100\,yr) before coalescence
\citep[see][]{baumgardt04b,zwart04}.


\acknowledgements

Special thanks go to J. Fregeau and S. Rappaport for providing
valuable critical remarks on the paper.  I would also like to
acknowledge enlightening discussions with R. Di Stefano, S. Gaudi,
A. Juett, A. Loeb, and S. Ransom.  This work was supported by NASA and
the Chandra Postdoctoral Fellowship program through grant number
PF2-30024.




\end{document}